\documentclass{PoS}
\usepackage{amsmath,amssymb,cite,epsfig,slashed,multicol,kotex}
\usepackage{tikz}
\usepackage[compat=1.1.0]{tikz-feynman}
\usetikzlibrary{shapes,arrows,positioning,automata,backgrounds,calc,er,patterns,trees,decorations.pathmorphing,decorations.markings}
\title{Unitarizing SIMP scenario with dark vector resonances}
\ShortTitle{Unitarizing SIMP scenario with dark vector resonances}
\author{\speaker{Soo-Min Choi}\\
        Department of Physics, Chung-Ang University, Seoul 06974, Korea\\
        E-mail: \email{soominchoi90@gmail.com}}
\author{Hyun Min Lee\\ 
        Department of Physics, Chung-Ang University, Seoul 06974, Korea\\
        School of Physics, Korea Institute for Advanced Study (KIAS), Seoul 02455, Korea\\
        E-mail: \email{hminlee@cau.ac.kr}}
\author{Pyungwon Ko\\ 
        School of Physics, Korea Institute for Advanced Study (KIAS), Seoul 02455, Korea\\
        E-mail: \email{pko@kias.re.kr}}
\author{Alexander Natale\\ 
        School of Physics, Korea Institute for Advanced Study (KIAS), Seoul 02455, Korea\\
        E-mail: \email{alexnatale@kias.re.kr}}                
\abstract{Dark pion is the one of the most natural and realistic model to make $3 \rightarrow 2$ contact interactions among models of the Strongly Interacting Massive Particle (SIMP). In this model, there is a perturbativity problem to satisfy the relic density with the self-annihilation. So we show the perturbativity problem can be alleviated and the SIMP pion model become more realistic by including dark vector mesons.}
\FullConference{ICHEP 2018, International conference on high energy physics\\
		COEX, Seoul, Republic of Korea.\\
		July, 4--11 2018}

\begin{document}
\section{Introduction}
\noindent A lot of indirect evidences for dark matter have been presented.  Among them, observation of the Bullet cluster shows the presence of dark matter from the gravitational lensing effect and the X-ray spectrum for visible gases after collision of two clusters and it gives an upper bound on the self-interaction of dark matter, $\sigma_{\rm self}/m_{\rm DM} \lesssim 1\ {\rm cm^2/g}$.\\
\indent Weakly Interacting Massive Particle (WIMP) is a candidate for dark matter that causes a "Miracle" and satisfies the relic density of dark matter with weak coupling and weak scale mass. Also, the self-interaction of WIMP is much smaller than the upper bound of the Bullet cluster. There are so many efforts to find WIMP in experiments, but we couldn't find a direct evidence of WIMP yet.\\
\indent In this situation, the small scale problems have required self-interaction close to the upper bound of the bullet cluster. The small scale problems mean differences between simulations and observations of $\Lambda$CDM. Because large self-interaction ($0.1\sim 1\ {\rm cm^2/g}$) can alleviate the small scale problem, researches on the self-interacting dark matter are very interesting.\\
\indent It is natural to have a large self-interaction for light dark matter. In particular, Strongly Interacting Massive Particle (SIMP) has recently drawn a lot of attention \cite{SIMP}. SIMP is a plausible candidate for dark matter that satisfies relic density through $3 \rightarrow 2$ self-annihilation.\\
\section{Pion SIMP with Vector Mesons}
\noindent The simplest realization of the $3\rightarrow 2$ interaction is definitely a 5-point contact interaction and the one of the best way to realize this contact interaction is Wess-Zumino-Witten (WZW) Lagrangian in the chiral perturbation theory (ChPT) \cite{WZW1}. This is a low energy dark pion model from the breaking of $SU(3)_L \times SU(3)_R \rightarrow SU(3)_V$ subgroup by the dark quark condensation. In this theory, WZW Lagrangian can be added by anomaly and it makes the 5 point interaction of the dark pions. However, minimal SIMP with pions only has perturbativity problem, so leading order ChPT could be broken down. It means that an expansion parameter of this theory exceeds the perturbativity bound to satisfy the relic density for the mass solving the small scale problem \cite{pert}.\\
\indent To alleviate the perturbativity problem and to make a more realistic model, we consider dark vector mesons in this theory \cite{SIMPVM}. Because the pion Lagrangian gotten from the non-linearly transformed coset field theory is same with the pion Lagrangian gotten by integrating out the gauge bosons after gauging this subgroup, the vector mesons can be included easily as massive gauge fields of a gauged local $SU(3)_V$ in the ChPT. Detailed discussion on the model can be found in Ref. \cite{SIMPVM}. For simplicity, we consider degenerate pion and vector meson masses. Relevant Lagrangian terms for the $3\rightarrow 2$ channel are 
%
\begin{equation}
\begin{aligned}
\mathcal{L} 
\supset & - i a g {\rm Tr}(V_\mu[\partial^\mu \pi, \pi])-\frac{a}{4f_\pi^2}{\rm Tr}([\pi, \partial_\mu \pi]^2) - \frac{2N_c}{15\pi^2 f_\pi^5}\epsilon^{\mu\nu\rho\sigma} {\rm Tr}(\pi \partial_\mu \pi\partial_\nu \pi\partial_\rho \pi\partial_\sigma \pi)\\
& -\frac{i g N_c(c_1-c_2)}{4\pi^2 f_\pi^3}\epsilon^{\mu\nu\rho\sigma}{\rm Tr}(V_\mu \partial_\nu \pi\partial_\rho \pi\partial_\sigma \pi ) + \frac{g N_c c_3}{8\pi^2 f_\pi}\epsilon^{\mu\nu\rho\sigma}{\rm Tr}(\partial_\mu V_\nu (V_\rho \partial_\sigma \pi - \partial_\rho \pi V_\sigma ))
\end{aligned}
\end{equation}
where $g$ is a gauge coupling, $f_\pi$ is a decay constant of $\pi$, $a = m_V^2/(g f_\pi)^2$, and $N_c$ is the number of colors. Then, there are 7 types of the 5 point interactions with 2 types of resonances.
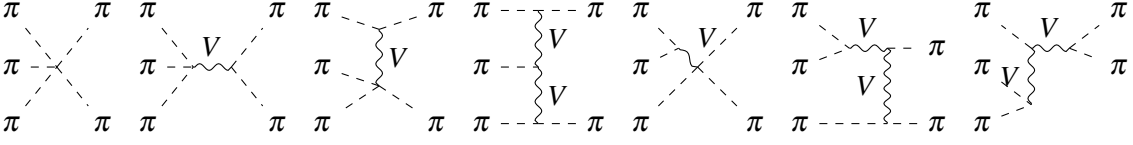
\begin{figure}[h!]
\begin{center}
\begin{tikzpicture}
\begin{feynman}
    \vertex (i1) {$\pi$};
    \vertex at ($(i1) + (0.0cm, -0.75cm)$) (i2) {$\pi$};
    \vertex at ($(i1) + (0.0cm, -1.5cm)$) (i3) {$\pi$};
    \vertex at ($(i1) + (1.2cm, +0.0cm)$) (f1) {$\pi$};
    \vertex at ($(i1) + (1.2cm, -1.5cm)$) (f2) {$\pi$};
    \vertex at ($(i1) + (0.6cm, -0.75cm)$) (m1);
    \diagram {
      {(i1),(i2),(i3)} -- [scalar] (m1) -- [scalar] {(f1),(f2)},    
    };
\end{feynman}
\end{tikzpicture}
\begin{tikzpicture}
\begin{feynman}
    \vertex (i1) {$\pi$};
    \vertex at ($(i1) + (0.0cm, -0.75cm)$) (i2) {$\pi$};
    \vertex at ($(i1) + (0.0cm, -1.5cm)$) (i3) {$\pi$};
    \vertex at ($(i1) + (1.7cm, +0.0cm)$) (f1) {$\pi$};
    \vertex at ($(i1) + (1.7cm, -1.5cm)$) (f2) {$\pi$};
    \vertex at ($(i1) + (0.6cm, -0.75cm)$) (m1);
    \vertex at ($(i1) + (1.1cm, -0.75cm)$) (m2);
    \diagram {
      {(i1),(i2),(i3)} -- [scalar] (m1) -- [photon, edge label = $V$] (m2) -- [scalar] {(f1),(f2)},      
    };
\end{feynman}
\end{tikzpicture}
\begin{tikzpicture}
\begin{feynman}
    \vertex (i1) {$\pi$};
    \vertex at ($(i1) + (0.0cm, -0.75cm)$) (i2) {$\pi$};
    \vertex at ($(i1) + (0.0cm, -1.5cm)$) (i3) {$\pi$};
    \vertex at ($(i1) + (1.5cm, +0.0cm)$) (f1) {$\pi$};
    \vertex at ($(i1) + (1.5cm, -1.5cm)$) (f2) {$\pi$};
    \vertex at ($(i1) + (0.75cm, -0.25cm)$) (m1);
    \vertex at ($(i1) + (0.75cm, -1.0cm)$) (m2);
    \diagram {
      {(i1),(f1)} -- [scalar] (m1) -- [photon, edge label = $V$] (m2) -- [scalar] {(i2),(i3),(f2)},      
    };
\end{feynman}
\end{tikzpicture}
\begin{tikzpicture}
\begin{feynman}
    \vertex (i1) {$\pi$};
    \vertex at ($(i1) + (0.0cm, -0.75cm)$) (i2) {$\pi$};
    \vertex at ($(i1) + (0.0cm, -1.5cm)$) (i3) {$\pi$};
    \vertex at ($(i1) + (1.5cm, +0.0cm)$) (f1) {$\pi$};
    \vertex at ($(i1) + (1.5cm, -1.5cm)$) (f2) {$\pi$};
    \vertex at ($(i1) + (0.75cm, -0.0cm)$) (m1);
    \vertex at ($(i1) + (0.75cm, -0.75cm)$) (m2);
    \vertex at ($(i1) + (0.75cm, -1.5cm)$) (m3);
    \diagram {
      {(i1),(f1)} -- [scalar] (m1) -- [photon, edge label = $V$] (m2),
      (i2) -- [scalar] (m2) -- [photon, edge label = $V$] (m3),
      {(i3),(f2)} -- [scalar] (m3),      
    };
\end{feynman}
\end{tikzpicture}
\begin{tikzpicture}
\begin{feynman}
    \vertex (i1) {$\pi$};
    \vertex at ($(i1) + (0.0cm, -0.75cm)$) (i2) {$\pi$};
    \vertex at ($(i1) + (0.0cm, -1.5cm)$) (i3) {$\pi$};
    \vertex at ($(i1) + (1.5cm, +0.0cm)$) (f1) {$\pi$};
    \vertex at ($(i1) + (1.5cm, -1.5cm)$) (f2) {$\pi$};
    \vertex at ($(i1) + (0.5cm, -0.5cm)$) (m1);
    \vertex at ($(i1) + (0.75cm, -0.75cm)$) (m2);
    \diagram {
      {(i1),(i2)} -- [scalar] (m1) -- [photon, edge label = $V$] (m2) -- [scalar] {(i3),(f1),(f2)},      
    };
\end{feynman}
\end{tikzpicture}
\begin{tikzpicture}
\begin{feynman}
    \vertex (i1) {$\pi$};
    \vertex at ($(i1) + (0.0cm, -0.75cm)$) (i2) {$\pi$};
    \vertex at ($(i1) + (0.0cm, -1.5cm)$) (i3) {$\pi$};
    \vertex at ($(i1) + (1.8cm, -0.5cm)$) (f1) {$\pi$};
    \vertex at ($(i1) + (1.8cm, -1.5cm)$) (f2) {$\pi$};
    \vertex at ($(i1) + (0.65cm, -0.5cm)$) (m1);
    \vertex at ($(i1) + (1.15cm, -0.5cm)$) (m2);
    \vertex at ($(i1) + (1.15cm, -1.5cm)$) (m3);
    \diagram {
      {(i1),(i2)} -- [scalar] (m1) -- [photon, edge label = $V$] (m2) -- [scalar] (f1),
      {(i3),(f2)} -- [scalar] (m3) -- [photon, edge label = $V$] (m2)      
    };
\end{feynman}
\end{tikzpicture}
\begin{tikzpicture}
\begin{feynman}
    \vertex (i1) {$\pi$};
    \vertex at ($(i1) + (0.0cm, -0.75cm)$) (i2) {$\pi$};
    \vertex at ($(i1) + (0.0cm, -1.5cm)$) (i3) {$\pi$};
    \vertex at ($(i1) + (1.8cm, -0.0cm)$) (f1) {$\pi$};
    \vertex at ($(i1) + (1.8cm, -0.75cm)$) (f2) {$\pi$};
    \vertex at ($(i1) + (0.65cm, -1.25cm)$) (m1);
    \vertex at ($(i1) + (0.65cm, -0.5cm)$) (m2);
    \vertex at ($(i1) + (1.15cm, -0.5cm)$) (m3);
    \diagram {
      {(i2),(i3)} -- [scalar] (m1) -- [photon, edge label = $V$] (m2) -- [scalar] (i1),
      (m2) -- [photon, edge label = $V$] (m3) -- [scalar] {(f1),(f2)}     
    };
\end{feynman}
\end{tikzpicture}
\end{center}
  \caption{7 types of the 5 point interactions with 2 types of resonances.}
  \label{Fig1}
\end{figure}
\\
\indent For example, let's consider 3-pion resonance case ($m_V = 3m_\pi\sqrt{1+\epsilon_V}$) to check how much vector mesons alleviate the perturbativity problem. This resonance example with narrow width approximation (NWA) shows the largest effect of vector meson contribution. Also, we set the parameters or couplings of the Lagrangian as SM parameters as possible to make a realistic model.\\
\begin{figure}[h!]
\centering
 \includegraphics[width=0.45\textwidth]{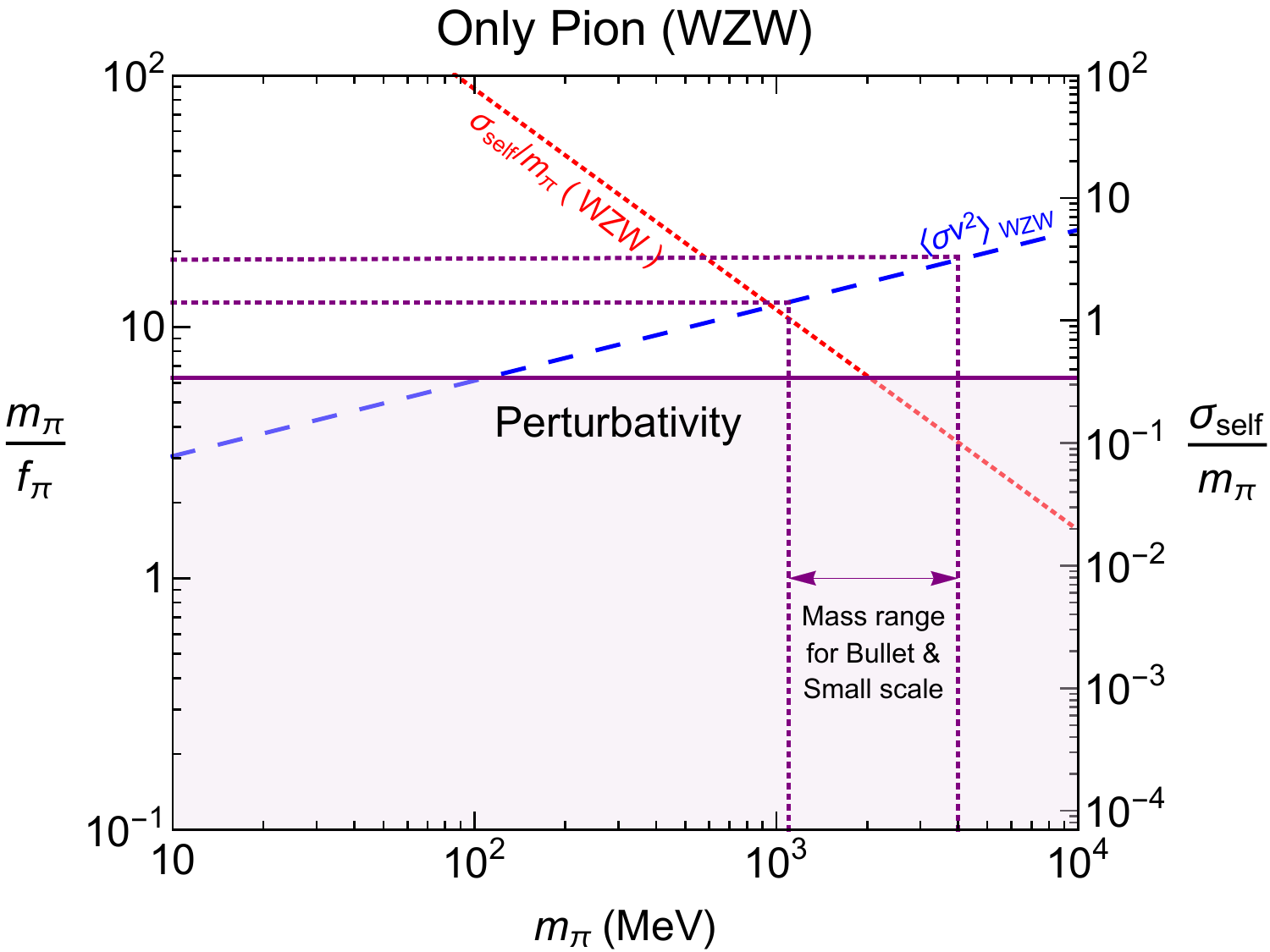} \ 
 \includegraphics[width=0.45\textwidth]{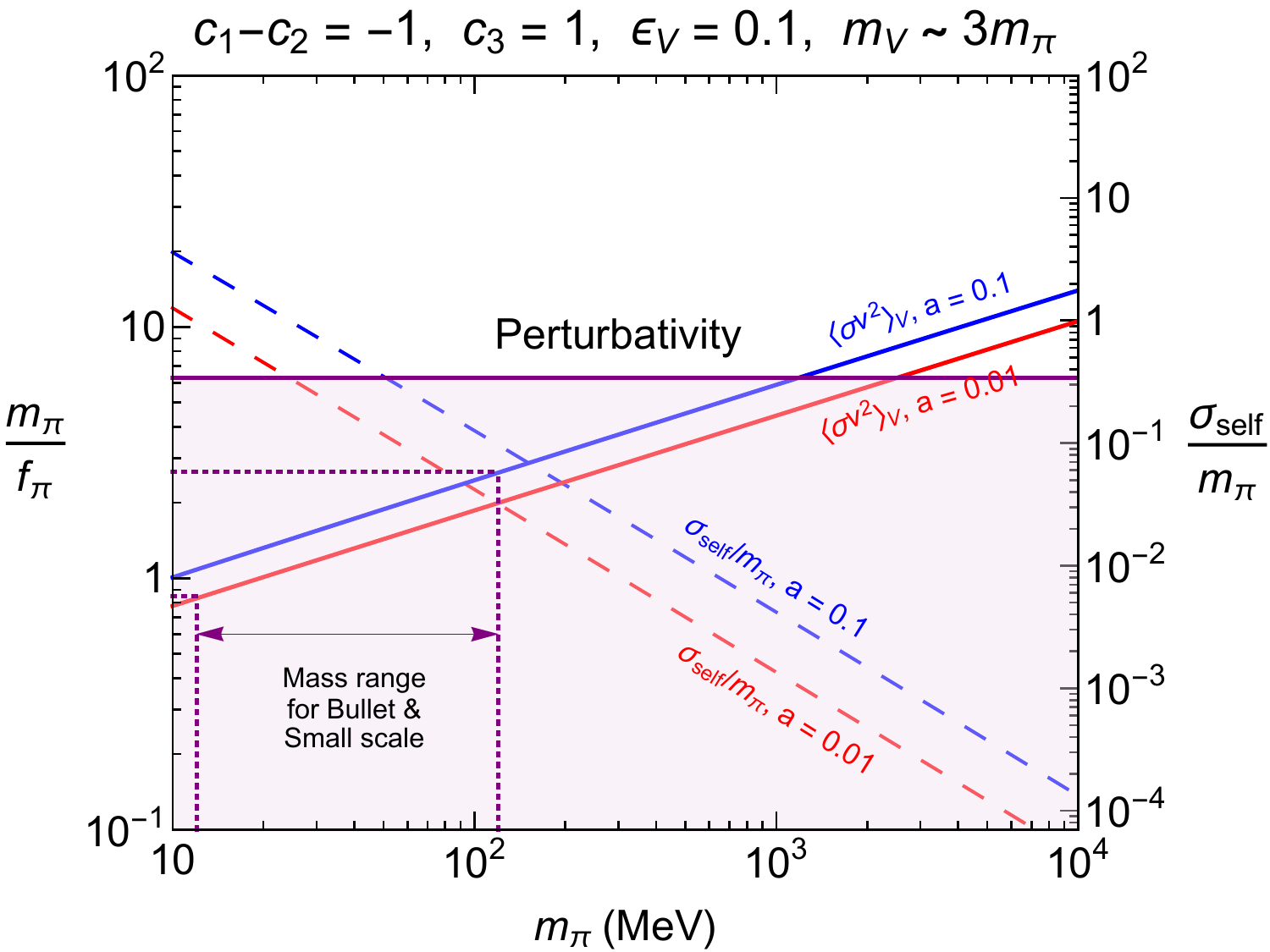}
  \caption{Left panel is the case of only WZW interaction and there is a perturbativity problem for the mass range of the Bullet cluster and small scale problems. Right panel show that expansion parameter $m_\pi/f_\pi$ doesn't exceed perturbativity bound by the vector meson mediated channels.}
\end{figure}
\\
\indent In the case of 3-pion resonance, 3-body decays of vector mesons to 3-pions are negligible, because the 3-body decay width is proportional to $\epsilon_V^2$ by phase space. So the decay width of vector mesons is almost decided by 2-pion decay of vector meson and this channel makes an upper bound of $a$ when NWA is used ($m_V\Gamma_V/(9 m_\pi^2)\lesssim 0.1$). As a result, the maximal value of $a$ is 0.1 near 3-pion resonance. \\
\indent Left panel of the figure 2 shows that the order parameter exceeds the perturbativity bound ($m_\pi/f_\pi \sim 2\pi$) for the pion SIMP with the only WZW model. Right panel shows that expansion parameter does not exceed perturbativity bound for the pion SIMP with the vector mesons. So leading order analysis of this model is believable. Also, parameter space is wider than the only WZW case. Moreover, in the mass range of light dark matter ($m_\pi \lesssim 1\ {\rm GeV}$), order parameter always satisfies perturbativity bound, $m_\pi/f_\pi < 6(4.5)$, for $a = 0.1(0.01)$. Lastly, there is an enough parameter space for the various values of this model parameters.
\section{Kinetic Equilibrium and General Masses}
\noindent Because the one of the pion masses goes to kinetic energy of dark matter in the $3\rightarrow 2$ channel, kinetic equilibrium condition with standard model particles is very important for SIMP \cite{SIMP}. There are possibilities for kinetic equilibrium if we include singlet scalar, sigma field or $Z'$ and mixing with the SM particles \cite{SIMPkin}. Singlet scalar or sigma field can get a mixing with SM Higgs. In that case, There are two ways, (1) pion-electron scattering and (2) pion-Higgs scattering with decaying of Higgs to SM fermions. Also $Z'$ can mediate pion-electron scattering for the kinetic equilibrium.\\
\indent Moreover, a study of general masses for pions and vector mesons is important. If we consider broad spectrum of vector meson masses, there are many 3(2)-pion resonance points for self-annihilation. It will give us rich parameter space than the case of degenerated masses. The second interesting possibility is multi-component SIMP pion. If the mass difference of two lightest pions are small, they could be considered as multi-component dark matters. So there are interesting sides to study for non-degenerated masses.
\section{Conclusions}
\noindent We have shown that dark pion SIMP in the dark ChPT can be dark matter candidates. Especially by including the vector mesons, $3\rightarrow 2$ cross-sections are enhanced and perturbativity problem of the expansion parameter can be alleviated in dark pion SIMP scenario. For a complete study of this model, kinetic equilibrium and more general masses are also very interesting.
\section{Acknowledgments}
\noindent The work of S. M. C. and H. M. L. are supported in part by Basic Science Research Program through the National Research Foundation of Korea Grant No. NRF- 2016R1A2B4008759. The work of S. M. C. is supported in part by TJ Park Science Fellowship of POSCO TJ Park Foundation. The work of P. K. and A. N. are supported in part by Korea Research Foundation Research Grant No. NRF-2015R1A2A1A05001869.
\end{document}